\documentclass[aps,twocolumn,showpacs,preprintnumbers,prl]{revtex4}
\usepackage{graphicx}

\newcommand{\R}{{\bf r}}
\newcommand{\RR}{{\bf r}'}
\newcommand{\RP}{{\bf r}''}
\newcommand{\PP}{{\bf p}}
\newcommand{\PPP}{{\bf p}'}

\newcommand{\UP}{n_{\uparrow}}
\newcommand{\DN}{n_{\downarrow}}
\newcommand{\TUP}{\tau_{\uparrow}}
\newcommand{\TDN}{\tau_{\downarrow}}

\newcommand{\be}{\begin{equation}}
\newcommand{\ee}{\end{equation}}
\newcommand{\bea}{\begin{eqnarray}}
\newcommand{\eea}{\end{eqnarray}}
\newcommand{\bean}{\begin{eqnarray*}}
\newcommand{\eean}{\end{eqnarray*}}

\begin{document}

%\title{High-Level Correlated Approach to the Jellium Surface Energy}
\title{Collapse of the Electron Gas to Two Dimensions in Density Functional Theory}
\author{Lucian A. Constantin$^1$, John P. Perdew$^1$, and J. M.
Pitarke$^{2,3}$}
\affiliation{
$^1$Department of Physics and
Quantum Theory Group, Tulane University, New Orleans, LA 70118\\
$^2$CIC nanoGUNE Consolider, Mikeletegi Pasealekua 56, E-20009 Donostia, Basque
Country\\
$^3$Materia Kondentsatuaren Fisika Saila, UPV/EHU, and Centro F\'\i
sica Materiales CSIC-UPV/EHU,\\
644 Posta kutxatila, E-48080 Bilbo, Basque Country}

\date{\today}

\begin{abstract}
Local and semilocal density-functional approximations for the
exchange-correlation energy fail badly in the zero-thickness limit of a
quasi-two-dimensional electron gas, 
where the density variation is rapid almost everywhere.
Here we show that a fully nonlocal fifth-rung functional, the inhomogeneous 
Singwi-Tosi-Land-Sj\"olander (STLS) approach, which employs both occupied and unoccupied 
Kohn-Sham orbitals, recovers
the true two-dimensional STLS limit and appears to be remarkably accurate for any thickness 
of the slab
(and thus for the dimensional crossover).  We also show that this good behavior is only 
partly due to the use of the full exact exchange energy.
\end{abstract}


\maketitle

One of the main goals of electronic-structure theory is prediction
of the energy of inhomogeneous interacting
many-electron systems.  
Time-independent and time-dependent density
functional theories (DFT and TDDFT)~\cite{X} provide calculable predictions
respectively for ground- and excited-states, and are
intimately linked together. In these theories, the noninteracting
kinetic energy is treated as an exact functional of the occupied 
Kohn-Sham (KS) orbitals~\cite{KS}, and only the exchange-correlation (xc)
energy $E_{xc}$ and/or potential $v_{xc}(\R)$
have to be approximated. These theories entail a
hierarchy of approximations for exchange and correlation.
The more sophisticated approximations satisfy many exact
conditions, but must still be carefully tested for accuracy
and reliability. In this paper, we test the 
inhomogeneous Singwi-Tosi-Land-Sj\"olander (STLS) method~\cite{DWG} for 
the quasi two-dimensional (2D) electron gas, a problem that becomes increasingly 
challenging for density functionals as the true 2D or zero-thickness limit 
is approached~\cite{2d,BPE}.

The ladder classification ~\cite{jacob} of ground-state density functionals 
for $E_{xc}$ has three complete non-empirical
rungs: the local-spin-density approximation
(LSDA)~\cite{PW}, the generalized-gradient approximation
(GGA)~\cite{PBE}, and the meta-GGA~\cite{TPSS}. The meta-GGA,
which satisfies many exact constraints but still uses to some extent
the error cancellation between exchange and correlation, has as ingredients
the spin densities $\UP$ and $\DN$, their gradients $\nabla\UP$ and
$\nabla\DN$, and the KS non-interacting kinetic energy densities 
$\TUP$ and $\TDN$. These local and semilocal density functionals 
(LSDA, GGA, and meta-GGA) work for atoms, molecules, solids, and 
surfaces~\cite{KPB}. They also work for atomic monolayers~\cite{RAM} 
and other quasi-2D systems~\cite{RM}, but they fail as the true 2D limit is
approached~\cite{PoP,KLNLHM}, because of the high inhomogeneity of the
electron density along the confined direction. The failure of local and 
semilocal density functionals to describe the dimensional crossover 
of the exact xc functional has been avoided by using nonlocal 
models such as the weighted-density approximation~\cite{GGG}.

The next rung of the ladder is the hyper-GGA (HGGA)~\cite{Per1}, a
nonlocal-functional approximation which uses the 
Tao-Perdew-Staroverov-Scuseria (TPSS) meta-GGA~\cite{TPSS}
ingredients and the conventional exact exchange energies per particle
$\epsilon_{x\uparrow}$ and $\epsilon_{x\downarrow}$~\cite{TSSP}. For the xc
energy per particle at position $\R$, one writes:
\begin{equation}
\epsilon_{xc}^{HGGA}=\epsilon_{x}+\left[1-a\right]
\left[\epsilon_{x}^{TPSS}-\tilde{\epsilon}_{x}\right]+
\epsilon_{c}^{TPSS}.
\label{el2}
\end{equation}
Here $\tilde{\epsilon}_{x}(\R)$ is the exact exchange energy per particle in the
TPSS gauge 
\cite{TSSP}, and $a(\R)$ is a nonlocal functional bounded between 0 and 1. The
second term on 
the 
right-hand side of Eq. (\ref{el2}) is the static correlation and was built
such that, 
together with the TPSS (dynamic) correlation, it is compatible
with exact exchange. The mixing parameter $a(\R)$ goes to 1 when exact exchange
is much bigger than 
correlation (for one-electron systems 
and for the high-density limit), 
when the density is rapidly varying, and when an open system has a high
fluctuation of electron number in spin-polarized regions at the Hartree-Fock
level.
HGGA satisfies more exact constraints than any semilocal functional. To balance
the full 
nonlocality of exact exchange, universal empirical parameters are invoked in
HGGA 
correlation and fitted to chemical data.
The set of parameters adopted here is for use with TPSS orbitals.
We show that this HGGA
partially avoids the bad behavior
of the semi-local xc density functionals for the quasi-2D
electron gas.  

The central equation of TDDFT linear response (at frequency $\omega$), in which
all the objects are functionals of the ground-state density, is a Dyson-like
equation for the density-response function $\chi({\bf r},{\bf r}';\omega)$
(in atomic units where
$e^2=\hbar=m_e=1$)~\cite{GDP}:
\begin{eqnarray}\label{e1}
&&\chi({\bf r},{\bf r}';\omega)=
\chi_0({\bf r},{\bf r}';\omega)+\int d\R_1\,d\R_2\,
\chi_0({\bf r},{\bf r}_1;\omega)\cr\cr
&\times& \left\{{1\over |{\bf r}_1-{\bf r}_2|}+
f_{xc}[n]({\bf r}_1,{\bf r}_2;\omega)\right\}\,
\chi({\bf r}_2,{\bf r}';\omega),
\end{eqnarray}
where $\chi_0({\bf r},{\bf r}';\omega)$ is the density-response function of
non-interacting 
KS electrons and is exactly expressible in terms of KS orbitals~\cite{GK}, and 
$f_{xc}[n]({\bf r},{\bf r}';\omega)$ is the dynamic xc kernel which 
must be approximated.
When $f_{xc}[n]({\bf r},{\bf r}';\omega)$ is taken to be zero, 
Eq.~(\ref{e1}) reduces to 
the screening equation of the time-dependent Hartree or random phase
approximation (RPA). The xc energy
can then be 
calculated using the adiabatic-connection fluctuation-dissipation
formula~\cite{HG}, which allows TDDFT to produce sophisticated
approximations to the ground-state xc energy.
In particular, the RPA 
has been evaluated for the quasi-2D electron gas~\cite{GGG,note1b} 
and it reaches the RPA 2D electron gas limit; however, the RPA xc energy per
particle $\epsilon^{RPA}_{xc}$ for the 2D uniform electron gas underestimates
the exact $\epsilon_{xc}$ by more than $20$ $\rm{mHa}/e^-$~\cite{GGG}. In
order to go beyond the RPA, various approximations have been constructed for
the xc kernel, but they have mainly been taken from the 3D uniform electron
gas~\cite{CP} and do not take the dimensional crossover into account.

The inhomogeneous STLS (ISTLS)~\cite{DWG} is, like the RPA it corrects, a
“fifth-rung
density functional” that employs both occupied and unoccupied Kohn-Sham
orbitals.
The main idea of ISTLS (and STLS \cite{STLS}) relies on the
truncation of the
Bogoliubov-Born-Green-Kirkwood-Yvon (BBGKY) hierarchy of the kinetic
equations~\cite{BBGKY}, by assuming that the two-particle dynamic
pair-distribution function $f^{(2)}(\R,\PP,\RR,\PPP;t)$ can be expressed in
terms of the single-particle distribution function $f(\R,\PP;t)$ as
\begin{equation}
f^{(2)}(\R,\PP,\RR,\PPP;t)=g(\R,\RR)\,f(\R,\PP;t),f(\RR,\PPP;t),
\label{e2}
\end{equation}
where $g(\R,\RR)$ is the static (and momentum-independent) equilibrium
pair-correlation function. 

Using the linearity and time-invariance of the truncated BBGKY equation, Dobson
{\it et al.}~\cite{DWG} found the following Dyson-like "screening" integral 
equation for the density-response function:
\begin{equation}
\chi(\R,\RR;\omega)=\chi_0(\R,\RR;\omega)+\int d\RP
Q(\R,\RP;\omega)\chi(\RP,\RR;\omega),
\label{e3}
\end{equation}
where
\begin{equation}
Q(\R,\RR;\omega)=-\int d\RP \mbox{\boldmath$\nu$}_0(\R,\RP;\omega)\cdot
g(\RP,\RR)\nabla_{\RP}\frac{1}{|\RP-\RR|}.
\label{e4}
\end{equation}
Here, $\mbox{\boldmath$\nu$}_0(\R,\RR;\omega)$ is a vector response function,
which satisfies the equation
\begin{equation}
\chi_0(\R,\RR;\omega)=\nabla_{\RR}\cdot\mbox{\boldmath$\nu$}_0(\R,\RR;\omega).
\label{e5}
\end{equation}
The equilibrium pair-correlation function
$g(\R,\RR)$ is obtained from the fluctuation-dissipation theorem
\cite{HG}:
\begin{equation}
g(\R,\RR)=1-\frac{1}{\pi n(\R)n(\RR)}\int^\infty_0du\;\chi(\R,\RR;iu)
-\frac{\delta(\R-\RR)}{n(\RR)}.
\label{e6}
\end{equation}
Equations~(\ref{e3})-(\ref{e6}) are solved selfconsistently, until a 
converged
solution is obtained. 
This ISTLS scheme yields the exact exchange energy, 
as does the TDDFT scheme described above
(see Eq.~(\ref{e1})), and only the
correlation energy is approximated.
ISTLS correlation is exact for all
one-electron densities.

For the 2D and 3D uniform electron gases, the STLS approach made a remarkably
accurate prediction of the correlation energy over a wide range of densities
(for 3D: $1\leq r_s\leq 20$ and for 2D: $0.5\leq r_s^{2D}\leq 16$), as
confirmed by QMC calculations (see Table I of Ref.~\cite{CPDGLP} and
references therein)~\cite{note2b}. The STLS approximation is also known
to yield reasonable ground-state energies for the 1D and 2D Hubbard models in
the half-filled antiferromagnetic states~\cite{HV}.
Furthermore, the ISTLS, which is a “high-level correlated approach” that
predicts (and does not use as input) the correlation energy of the 3D and 2D
uniform gases, has been shown to yield accurate jellium xc surface energies
that are
close to their LSDA and RPA counterparts, and has been used to demonstrate
that a local-density approximation for the particle-hole interaction is
adequate to describe the surface energy of simple metals~\cite{CPDGLP}. 

A 2D uniform electron gas is described by the 2D electron-density parameter
$r_s^{2D}=1/\sqrt{\pi n^{2D}}=\sqrt{2}/k_F^{2D}$, where $n^{2D}$ is the density
of electrons per unit area, and $k_F^{2D}$ represents the magnitude of the
corresponding 2D Fermi wavevector. The 2D exchange energy per particle is
\begin{equation}
\epsilon^{2D}_x=-(4\sqrt{2}/(3\pi))/r_s^{2D}.
\label{ex}
\end{equation}
A realistic interpolation (which uses QMC data) between the high- and
low-density limits of the 2D correlation energy per particle is~\cite{PoP,SPL} 
\begin{equation}
\epsilon^{2D}_c=0.5058\left[\frac{1.3311}{(r^{2D}_s)^2}\left(\sqrt{1+1.5026
r^{2D}_s}-1\right)-\frac{1}{r^{2D}_s}\right].
\label{e12}
\end{equation}

For the description of a quasi-2D electron gas, we consider a quantum well of
thickness $L$ in the $z$-direction. In the infinite barrier model
(IBM)~\cite{Ne} for a quantum well, the KS
effective one-electron potential is replaced by zero inside infinitely high
potential walls at $z=0$ and $z=L$. Hence, the normalized KS one-electron wave
functions and energies at $0\leq z\leq L$ are
\begin{equation}
\phi_{l,{\bf k}_\parallel}({\bf r}_\parallel,z)=\sqrt{\frac{2}{AL}}\,
\sin\left(\frac{l\pi}{L}z\right)\,e^{i\bf{k}_{||}\cdot\bf{r}_{||}}
\label{e8}
\end{equation}
and
\begin{equation}
E_{l,k_\parallel}=\frac{1}{2}
\left[\left(\frac{l\pi}{L}\right)^2+k_{||}^2\right],
\label{e9}
\end{equation}
where $A$ represents the normalization area in the $xy$-plane, $l$ is the
subband index, and $\bf{r}_{||}$ and $\bf{k}_{||}$ represent the position and
the wavevector in the $xy$ plane. In this model the electrons cannot leak out
of the well, so the true 2D electron-gas limit is 
recovered by simply shrinking the well.
When only the lowest level is occupied ($E_{1,k_F^{2D}}<E_{2,0}$,
i. e., $L<\sqrt{3/2}\pi r_s^{2D}=L_{\rm max}$~\cite{PoP}), the density of
states of this quantum well begins to resemble the density of states of a 2D
electron gas, the motion in the $z$-direction is frozen out, and the system
can be considered quasi-two-dimensional. 

We would like to contract or expand the electron density $n(z)$ without
changing the total number of electrons per unit area. Hence, we perform a
one-dimensional scaling of the form $n_\lambda(z)=\lambda \,n(\lambda
z)$~\cite{LO}, which as $\lambda\to\infty$ yields the true 2D limit. This
scaled electron density coincides with the electron density that one would
find from Eq.~(\ref{e8}) by simply replacing the quantum-well thickness $L$ by
$L/\lambda$~\cite{PoP}. The corresponding exchange and correlation energies
per particle, $\varepsilon_x[n_\lambda]$ and $\varepsilon_c[n_\lambda]$,
should satisfy the following scaling relations~\cite{PoP}:
\begin{equation}
\lim_{\lambda\rightarrow\infty}\varepsilon_x[n_{\lambda}]>-\infty\;\;;\;\;\;
0>\lim_{\lambda\rightarrow\infty}\varepsilon_c[n_{\lambda}]>-\infty.
\label{e11}
\end{equation}
These equations, which start from those of Ref.~\cite{LO}, are not satisfied
by the LSDA, GGA, and meta-GGA.

Here we investigate the performance of the ISTLS approach to describe the IBM
quasi-2D electron gas for thicknesses $L$ such that $L<L_{\rm{max}}$.
We choose quasi-2D electron gases of fixed 2D electron-density parameters:
$r_s^{2D}=2/\sqrt{3}$ (as in Fig.~1 of Ref.~\onlinecite{GGG}) and
$r_s^{2D}=4$ (as in Figs.~2 and 3 of Ref.~\onlinecite{PoP}). The
self-consistent ISTLS scheme [Eqs.~(\ref{e3})-(\ref{e6})] was solved
numerically using occupied and unoccupied orbitals of the form of
Eq.~(\ref{e8}). Our numerical scheme is similar to the one described in
Ref.~\cite{CPDGLP};
however, in contrast to the jellium surface-energy calculation reported in
Ref.~\cite{CPDGLP}, where the ISTLS self-consistent scheme converges rapidly,
in the case of the quasi-2D electron-gas calculation the convergence is slow
(especially when $L<<L_{\rm{max}}$)~\cite{note2} and special care must be
taken to perform the frequency integration of Eq.~(\ref{e6}).

%%%%%%%%%%%%%%%%%%%%%%%%%%%%%%%%%%%%%%%%%%%%%%%%%%%%%
\begin{figure}
\includegraphics[width=\columnwidth]{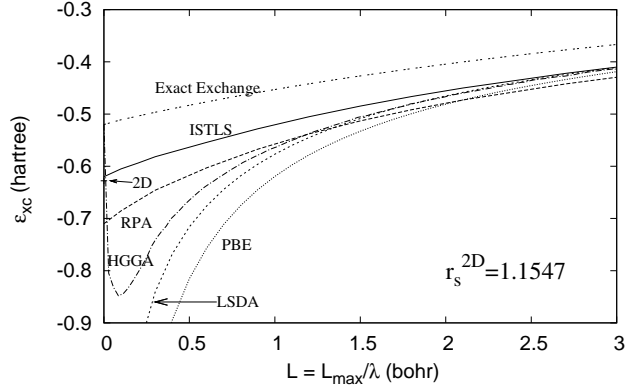}
\caption{Exchange-correlation energy per particle of an IBM quasi-2D electron
gas of fixed 2D electron density ($r_s^{2D}=2/\sqrt{3}$), as a function of the
quantum-well thickness $L_{\rm{max}}/\lambda$ ($L_{\rm{max}}=4.44$). The 2D
limit is from Eqs.~(\ref{ex})-(\ref{e12}). Various density-functional
approximations have been used
(LSDA, PBE-GGA, and HGGA), as well as the fifth-rung RPA and ISTLS. The exact
exchange energy per particle, $\varepsilon_x$, is also plotted, for
comparison. While the local and semilocal density functionals diverge
in the 2D limit, RPA and ISTLS calculations nicely recover the corresponding
xc energy of a 2D electron gas. The HGGA recovers the 2D exact exchange
limit; i.e., as in the case of the LSDA and the PBE-GGA, the
HGGA correlation energy per particle goes to zero in the 2D limit.}
\label{f2}
\end{figure}
%%%%%%%%%%%%%%%%%%%%%%%%%%%%%%%%%%%%%%%%%%%%%%%%%%%%%%%

%%%%%%%%%%%%%%%%%%%%%%%%%%%%%%%%%%%%%%%%%%%%%%%%%%%%%
\begin{figure}
\includegraphics[width=\columnwidth]{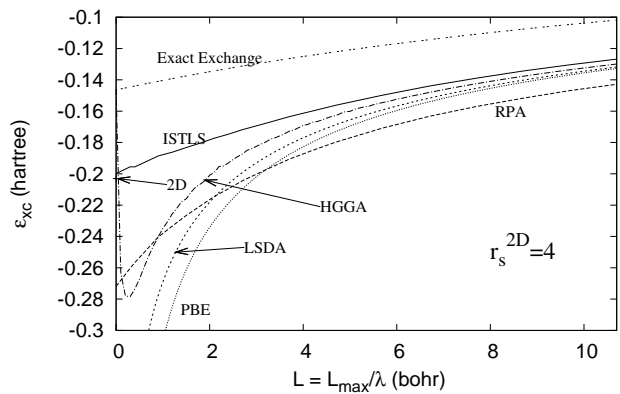}
\caption{As in Fig.~\ref{f2}, but now for $r_s^{2D}=4$ ($L_{\rm{max}}=15.39$).}
\label{f4}
\end{figure}
%%%%%%%%%%%%%%%%%%%%%%%%%%%%%%%%%%%%%%%%%%%%%%%%%%%%%%%

Figures~\ref{f2} and \ref{f4} show the results we have obtained, as a function
of the quantum-well thickness $L_{\rm max}/\lambda$, for the exact exchange
energy per particle, $\varepsilon_x$, and the xc energy per particle,
$\varepsilon_{xc}$, in the following approximations: LSDA, PBE-GGA, HGGA, RPA,
and ISTLS. We
observe that, while local and semilocal density-functional approximations (LSDA
and PBE-GGA) fail badly in the zero-thickness (2D) limit, both the RPA and the
ISTLS nicely recover their 2D counterparts, which in the case of the ISTLS
(-0.62 Hartree for $r_s^{2D}=2/\sqrt{3}$ and -0.20 Hartree for
$r_s^{2D}=4$~\cite{Jo}) are very close to the prediction of
Eqs.~(\ref{ex})-(\ref{e12})
or, equivalently, QMC calculations (-0.63
and -0.21, respectively). The RPA xc energy per particle, however,
considerably understimates $\varepsilon_{xc}$ for {\it all} slab
thicknesses~\cite{note3}. Figures~\ref{f2} and \ref{f4} show that the ISTLS xc
energy not only approaches closely the exact 2D limit but has also the correct
behavior in the limit 
$L\to L_{\rm{max}}$, which is expected to be well described
within the LSDA, GGA, and HGGA. The meta-GGA curves, not shown in
Figs.~\ref{f2} and \ref{f4}, are found to
be very close to their GGA counterparts. 
We note that the HGGA greatly improves over GGA
(and meta-GGA) and is more accurate than RPA for
almost all values of the slab thickness.

Most of the functionals tested in Figs.~\ref{f2} and \ref{f4} have been (exact
exchange, LSDA, PBE) or could be (HGGA, ISTLS) useful in condensed-matter
physics and quantum chemistry. 
For all of those except LSDA and ISTLS,
the correlation
energy per electron tends to zero as the quantum-well thickness $L_{\rm
max}/\lambda$ goes to zero; for ISTLS, it tends to a realistic negative value.

In summary, we have shown that the ISTLS approach~\cite{DWG} correctly and remarkably
describes the dimensional crossover (from 3D to 2D) of the xc energy. The quasi-2D electron gas is an
important and difficult test for density-functional approximations. This test
is related to the one-dimensional scaling, an exact constraint which is not
satisfied by LSDA, GGA, or meta-GGA.
The fourth-rung hyper-GGA with full exact exchange (see Eq.~(\ref{el2}))
is found to improve considerably the behavior of local
and semilocal density functionals over the whole thickness-range of the
quasi-2D electron gas. The fifth-rung ISTLS scheme, which uses as input all
occupied and unoccupied KS orbitals and is numerically more expensive~\cite{time}, is
found to be 
remarkably accurate for the description of various quasi-2D systems.
Due to its
accuracy in describing both the jellium
surface energy~\cite{CPDGLP} and the
energy per particle of a quasi-2D electron gas, we believe that further work
in testing and improving the ISTLS scheme could be important.  
So far, the computational cost of ISTLS has limited its applications to
situations in which the density variation is effectively
one-dimensional~\cite{CPDGLP}. Some
chemical tests have been carried out, however, at the RPA level~\cite{Furche},
and these calculations could well be extended within the full ISTLS
scheme.

L.A.C. and J.P.P. acknowledge NSF support (Grant No. DMR05-01588), 
and J.M.P. acknowledges support by the Spanish MEC and the EC NANOQUANTA. 
L.A.C. thanks the Donostia International Physics Center (DIPC), 
where this work was started.


\begin{thebibliography}{[Vo]}
\bibitem{2d} Physical systems with strong 2D character are found \cite{BPE}, e.g., in 
the widely used semiconductor heterojunctions, in electrons bound to the surface of
liquid helium, and in the copper-oxide planes 
of high-temperature superconductors.
\bibitem{note1b} In a variety of metal surfaces, the partially occupied band of
Shockley surface states with energies near the Fermi level forms a quasi-2D
electron gas that is immersed in a semi-infinite 3D gas of valence electrons.
This system was investigated in Ref.~\cite{PNSCZE} using the RPA.
\bibitem{note2b} The 3D parameter $r_s$ is the radius of a
sphere containing on average one electron; $r_s^{2D}$ represents the
radius of a circle that encloses on average one electron.
\bibitem{note2}
%four or five cycles yield a converged solution when $L\approx L_{\rm{max}}$.
For $L>L_{\rm{max}}$, four or five cycles yield a converged solution. For
$L<<L_{\rm{max}}$, however, ten or eleven cycles are needed in order
to obtain well-converged results.
\bibitem{note3} The RPA is known to yield both 2D and 3D 
correlation energies that are too negative. In the case of a slab, the correlation
energy of both the slab and the corresponding bulk material 
are too negative, yielding surface energies that are very close to more refined
calculations~\cite{CPDGLP}.
\bibitem{time} The computational time for the present 
calculations is less than one minute per $L$ value for LSDA and PBE, 
and less than ten minutes for HGGA, but can be
more than six hours for ISTLS.
\end{thebibliography}
\end{document}